\documentclass[preprint,showpacs]{revtex4}
\usepackage{amssymb,amsmath}
\usepackage[dvips]{graphicx}

\usepackage{dcolumn,epsfig}

\begin{document}

\title{Experimental investigation of water jets under gravity}
\author{Wellstandfree K. Bani, and Mangal C. Mahato}
\email{mangal@nehu.ac.in}
\affiliation{Department of Physics, North-Eastern Hill University,
Shillong-793022, India}

\begin{abstract}

The Plateau-Rayleigh theory essentially explains the breakup of liquid jets
as due to growing perturbations along the length of the jet. The essential idea
is supported by several experiments carried out in the past. Recently, the 
existence of a feedback mechanism in the form of recoil capillary waves was 
proposed to enhance the effect of the perturbations. We experimentally verify 
the existence of such recoil capillary waves. Using our experimental setup we 
further show that the wavy nature of the jet surface appears almost right after 
the emergence of the jet from the nozzle irrespective of the recoil 
capillary wave feedback. Moreover, our experimental results indicate existence of
a sharp boundary, along the length of the continuous jet, beyond which 
gravitational effect dominates over the surface tension.

\end{abstract}

\vspace{0.5cm}
\date{\today}


\maketitle

\section{Introduction}
There have been many attempts theoretically as 
well as experimentally to understand the physical mechanism behind the 
breaking up of continuous jets into drops. However, the study of this 
macroscopic phenomenon  continues to be of current 
interest\cite{Eggers1,Eggers2,Lin,Lautrup}.

The problem of instability of 
liquid jets was investigated by Plateau and then by Rayleigh and 
developed a theory which came to be known as Plateau-Rayleigh theory.
Rayleigh\cite{Ray1,Ray2}, based on surface energy considerations of inviscid 
liquids, showed that perturbations of wavelengths $\lambda$ larger than $\pi$ 
times the jet diameter $d_1$ grow rapidly with time. However, it is the fastest
growing perturbation ($\lambda\approx 4.508\times d_1$) that ultimately makes 
the jet column unstable against formation of droplets.
Chandrasekhar\cite{Chandra} later extended the theory to viscous liquids. Many 
experimental investigations have been conducted to examine the validity of 
Plateau-Rayleigh theory. The experiment of Goedde and Yuen, for example, 
applied external perturbations to study the length of the liquid jet before it 
breaks up\cite{Goedde}.

In some recent works,  Umemura and 
co-workers\cite{Umemura1,Umemura2,Umemura3,Umemura4} emphasized the 
idea that soon after the jet breaks up the new tip of the remaining column 
contracts to make its shape round once again to minimize its surface energy. 
The tip contraction (recoil) gives rise to upstream propagating capillary 
waves which upon reflection at the mouth of the nozzle move downstream with 
Doppler modified wavelengths. These feedback perturbations superpose with the 
preexisting perturbations and move down along the jet as combined 
perturbations. Some of these combined perturbations with the right wavelength 
cause the liquid column to breakup producing another contraction of the tip of 
the column and so on. 

\section{Experimental set up and experiment}
We set up and conduct an experiment to verify the existence and effect of the 
recoil capillary waves on the length of continuous water jet. We achieve  this
by damping the recoil capillary waves by bringing the jet in contact with a 
liquid surface beneath it. Moreover, when the continuous water jet smoothly
merges into the water it creates ripples on the water surface in the beaker.
Surface waves are observed using photographic methods and infer about the 
surface shape profile of the jet all along its length.

\begin{figure}[htp]
\centering
\includegraphics[width=13cm,height=8cm]{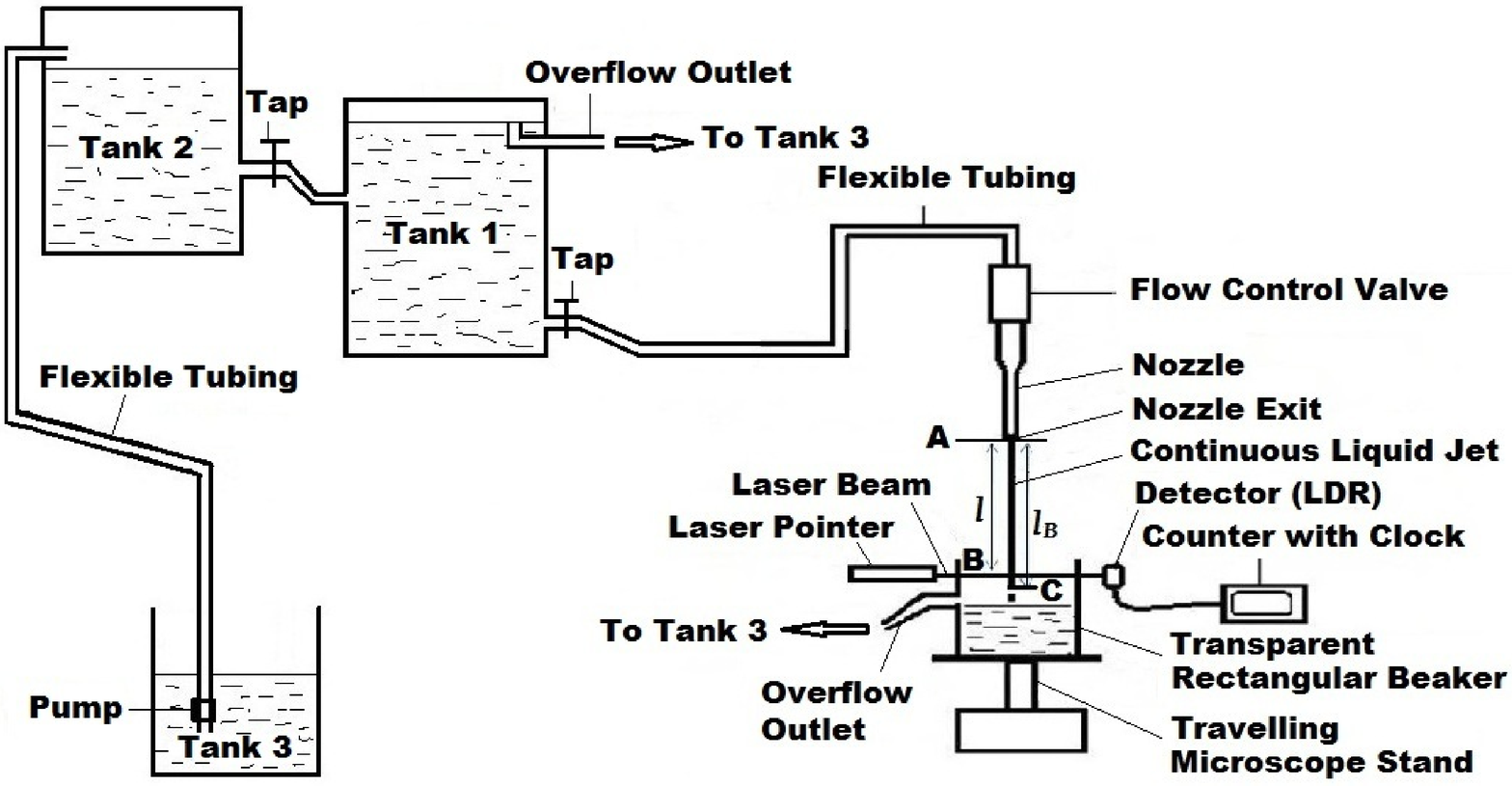}
\caption{A sketch of the experimental set up.}
\end{figure}

Our experimental set up is similar in essentials to that of Goedde and 
Yuen\cite{Goedde}. The new and important addition
is the intervening water containing beaker, Fig. 1. The details are given in 
Ref.\cite{Bani}. The transparent rectangular beaker with a level outlet on 
one of its vertical sides is placed vertically below the nozzle so that the 
water jet falls directly on (or smoothly merges into) the water kept in the 
beaker. The water level in the beaker is maintained fixed by letting the 
excess water flow out through the level outlet. The beaker is placed on a
horizontal platform fitted to the vertical stand of a travelling microscope 
(vernier scale least count = 0.001 cm) so that the beaker can be smoothly moved
vertically and its position measured. A  vertical-height adjustable 
laser-pointer-and-detector arrangement is also fitted to the platform so that 
the horizontal laser beam is incident normally on the vertical surface of the 
beaker and passes through the path of the water jet and then through the 
opposite surface of the beaker before it is collected by the detector.
A digital counter\cite{Details} with a clock is connected to the detector to count the 
number of discontinuities in the water jet over a period of time. 

Distilled water is issued vertically downward through a long glass nozzle (of 
length larger than about ten times its internal diameter $d$) in to the water 
in the beaker. The water flow rate is measured manually by collecting the jet 
water on a measuring cylinder for two minutes and calculating the mean value. 
As long as the jet remains continuous, at the level of the laser beam, the 
detector remains quiescent. However, a discontinuity in the jet after breakup 
allows the laser beam to pass through unobstructed and detected as a water drop 
count. We call the vertical distance between the mouth of the nozzle and the 
position of the laser beam as the jet-length, $l$. The vertical distance 
between the nozzle exit and the position of breakup of the jet, as detected by 
the laser-beam-counter, gives the breakup length, $l=l_B$.

Initially, the laser beam is made to face the continuous jet by moving the 
platform up closer to the nozzle, $l\approx 0$, and then the platform is 
gradually lowered in small steps so that $l$ increases. For each value of $l$, 
the number of counts is recorded for two minutes each for several times and 
their average calculated to obtain the mean drop-count rate. Naturally, 
the count rate begins from zero (at a threshold value of $l$) and then 
gradually keeps increasing as $l$ is increased in small steps. The jet-length 
$l$ at the very threshold point is termed here as the first breakup length 
$l_{FB}$ of the jet. The process is continued (by gradually increasing $l$) 
till the count rate reaches a saturation value. 

Throughout the above process the water flow rate is kept fixed. The same 
process is then repeated for several values of flow rates. Note that after 
each change of flow rate, the flow and the jet are allowed to become steady 
before the measurement process is begun. The same experiment is repeated for 
nozzles of various internal diameters $d$.
  
For our purpose, we perform two distinct sets of experiments. In the first set
(set 1), by adjusting the height of the laser beam arrangement, we let the 
laser beam pass just about 0.2 cm above the water surface on the beaker. In the
other set (set 2) the beam is kept at a height of about 1.5 cm above the water 
surface. Note that for the same position of the beaker, $l=l_1$ for the first 
set is larger by 1.3 cm than $l=l_2$ for the second set of experiments. 
Crucially, as explained below, the first breakup lengths $l_{FB}$ need not be 
the same for both the sets. 

Consider a situation wherein the jet begins to break up just about 2 mm 
above the water surface on the beaker. In the first set of experiments the 
counts just begin, that is, $l_1=l_{FB}$. However, if the moving tip of the 
remaining continuous jet touches the water surface before it gets the chance 
to recoil, the recoil capillary wave will get damped. On the other hand, 
consider a situation wherein the jet begins to breakup at about 1.5 cm above 
the water surface. This is the threshold point for the second set of 
experments, that is $l_2=l_{FB}$. In this case the tip of the remaining jet 
will have ample opportunity to recoil before it touches the water surface and 
hence recoil capillary waves will propagate up the jet undamped. The same will 
be the case even for the first set of experiments if the breakup were to take 
place at a somewhat larger height than 2 mm. Therefore, if the effect of recoil
capillary waves on the jet breakup length is to be a reality, $l_{FB}$ for the 
two sets of experiment must be different but the mean values of $l_B$ should 
be the same in both the sets of experiments.

Next, we observed the effect of water jet merging smoothly into the water in 
the beaker with the help of an ordinary (Nikon D5300) camera. We call the 
vertical distance between the mouth of the nozzle and the point at which the 
continuous jet touches the water surface again as jet-length but denote by the 
upper case $L$. We have taken photographs of the waves created on the 
water surface keeping the water surface at various positions (values of $L$) 
with respect to the stationary nozzle. From a submerged position of the nozzle,
the beaker arrangement was gradually lowered in stages and photographs taken. 
The photographs at various $L$ values show the nature of 
surface waves travelling towards the walls of the beaker. We measured the 
wavelengths of the waves (that is the mean separation between the successive 
crests of the waves) using the digitally stored photographs.

\section{Experimental results}

\begin{figure}
\centering
\includegraphics[width=12cm,height=8cm]{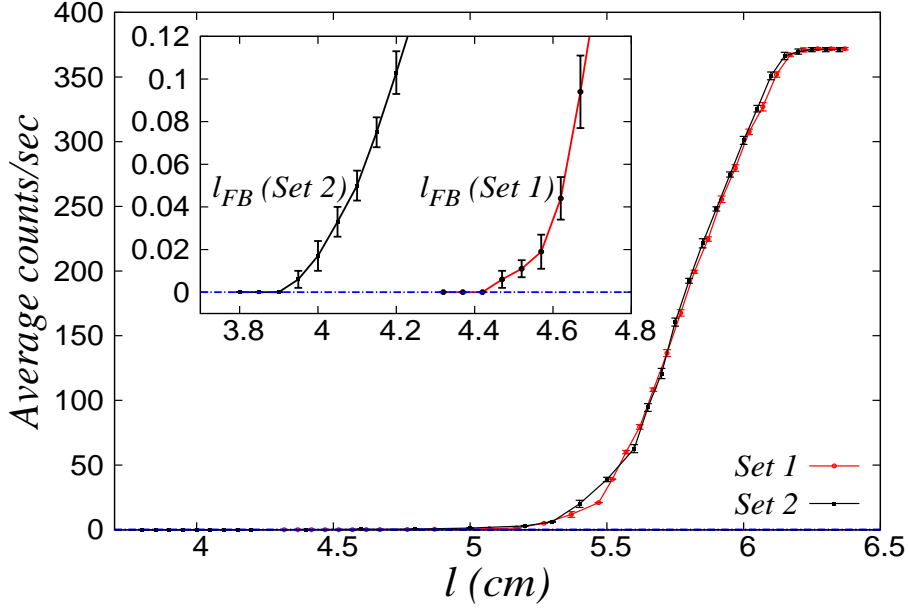}
\caption{Average count rate (s$^{-1}$) as a function of $l$ at the 
flow rate of 35.0 cc/min and $d=0.78$ mm for the two sets of experiments. The 
magnified graph (inset) show the first (jet) breakup points in the two
sets of experiments.}
\end{figure}

\begin{figure}
\centering
\includegraphics[width=12cm,height=8cm]{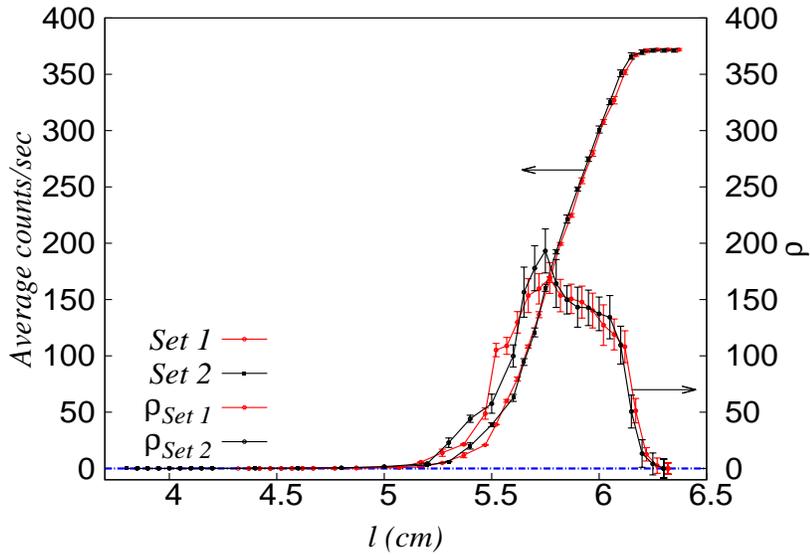}
\caption{Average count rate (s$^{-1}$) and probability distribution as a function of $l$ at the
flow rate of 35.0 cc/min and $d=0.78$ mm for the two sets of experiments.
The distribution, $\rho$, of breakup length is calculated
as the derivative of the count rate with respect to $l$ and the
normalized distribution $\rho_n$ = $\rho$/100.}
\end{figure}

\begin{figure}
\centering
\includegraphics[width=12cm,height=8cm]{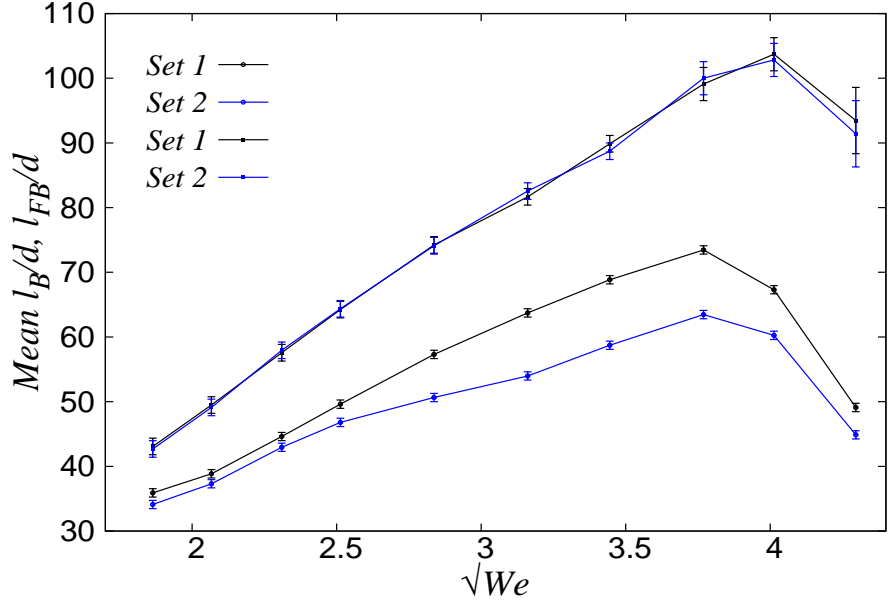}
\caption{The first breakup length $l_{FB}/d$ (lower set of two curves) and the 
mean breakup length $l_B/d$ (upper set of two curves) as a function of 
$\sqrt{We}$ for $d=0.78$ mm.}
\end{figure}

\begin{figure}
\centering
\includegraphics[width=12cm,height=8cm]{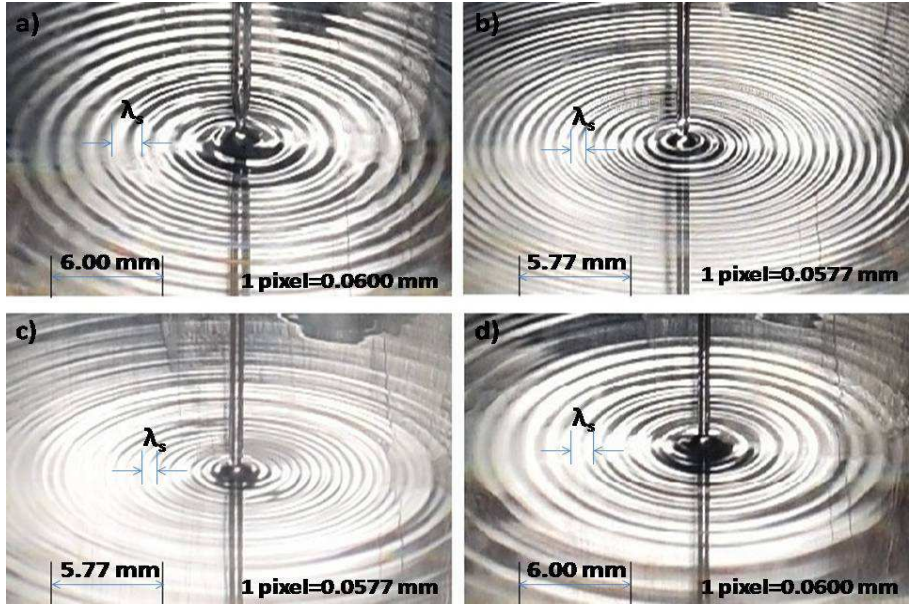}
\caption{Photograph of the water surface in the beaker for $d=0.95$ mm and
water flow rate of 40.0 cc/min. When the jet-length a)$L=0.168$ cm;
 b) $L=0.568$ cm; c) $L=3.668$ cm; and d) $L=4.268$ cm. The average wavelength 
($\lambda_s$) on the water surface just appear at a) then decrease with $L$
b) and c) and then increase till it reaches d) (i.e., just before the jet breaks into droplet).}
\end{figure}

\begin{figure}
\centering
\includegraphics[width=12cm,height=8cm]{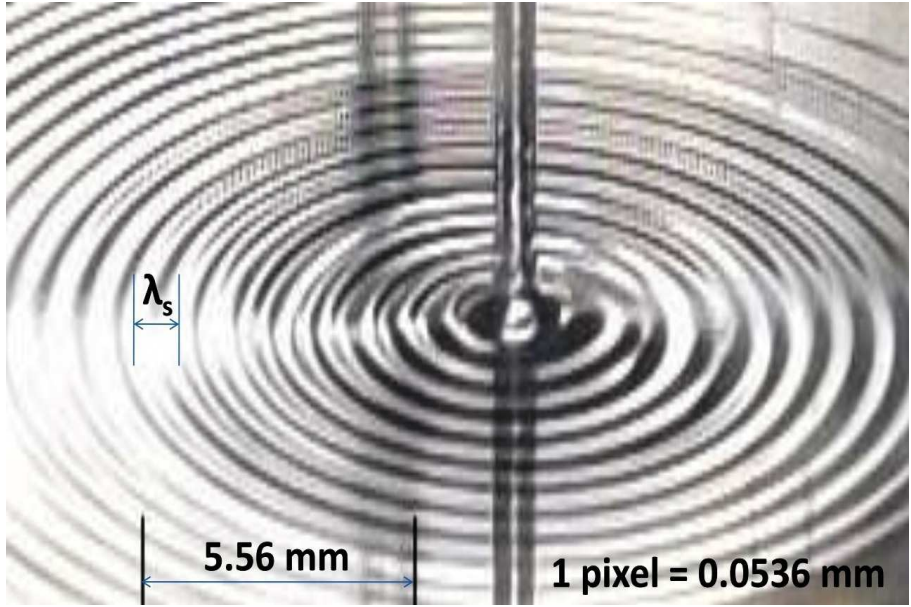}
\caption{Photograph of the water surface in the beaker when the jet-length
$L=0.942$ cm for $d=0.95$ mm and water flow rate of 50.0 cc/min.}
\end{figure}

\begin{figure}
\centering
\includegraphics[width=12cm,height=8cm]{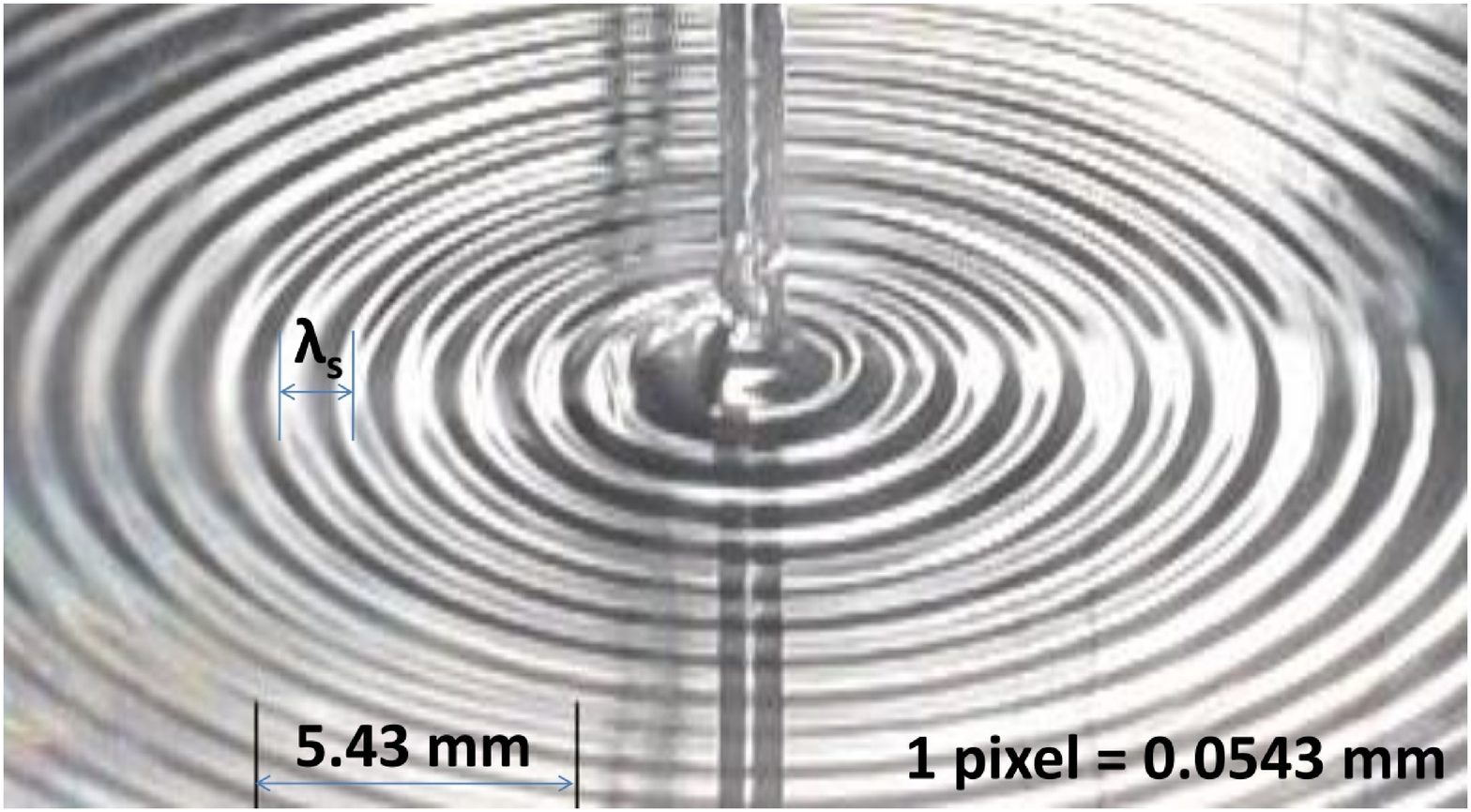}
\caption{Photograph of the water surface in the beaker when the jet-length
$L=1.295$ cm for $d=1.26$ mm and water flow rate of 72.5 cc/min.}
\end{figure}

\begin{figure}
\centering
\includegraphics[width=12cm,height=8cm]{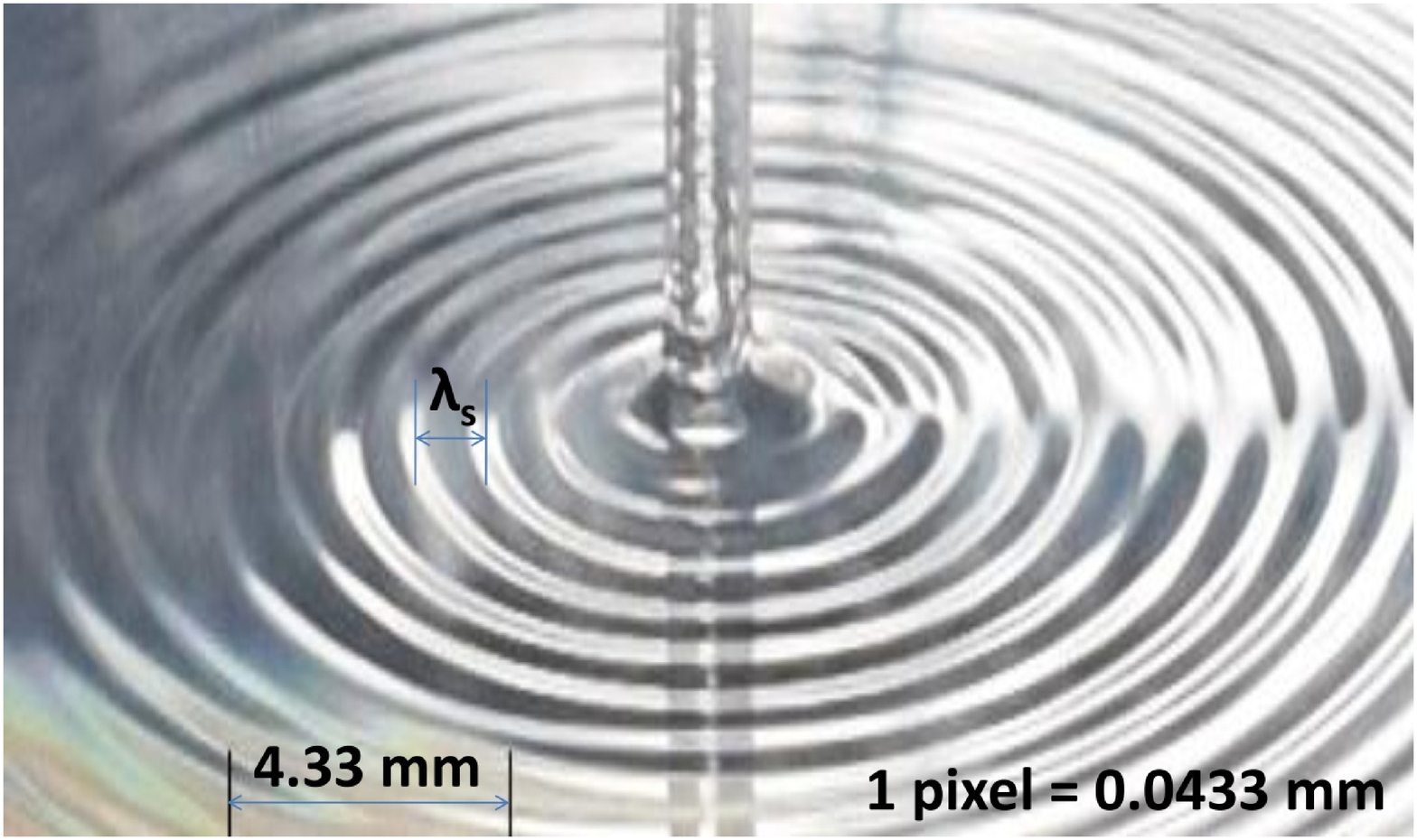}
\caption{Photograph of the water surface in the beaker when the jet-length
$L=1.806$ cm for $d=1.54$ mm and water flow rate of 120.0 cc/min.}
\end{figure}

All the measurements are done at the temperature of (25$\pm$.5)$^{\circ}$C and 
at relative humidity of (80$\pm$4)\%. However, in order to calculate the
Reynolds number Re ($=\frac{\rho_wud}{\mu}$) and the Weber number
We ($=\frac{\rho_wu^2}{\sigma}\frac{d}{2}$) we have used the tabulated values 
of surface tension $\sigma=72\times10^{-3}$ Nm$^{-1}$,  coefficient of dynamic 
viscosity $\mu=8.9\times 10^{-4}$ kgm$^{-1}$s$^{-1}$ and density 
$\rho_w=997.05$ kgm$^{-3}$ of water. The issuing jet speed $u$ is calculated 
as the ratio of the flow rate and the inner area of the nozzle exit.

\begin{figure}
\centering
\includegraphics[width=12cm,height=8cm]{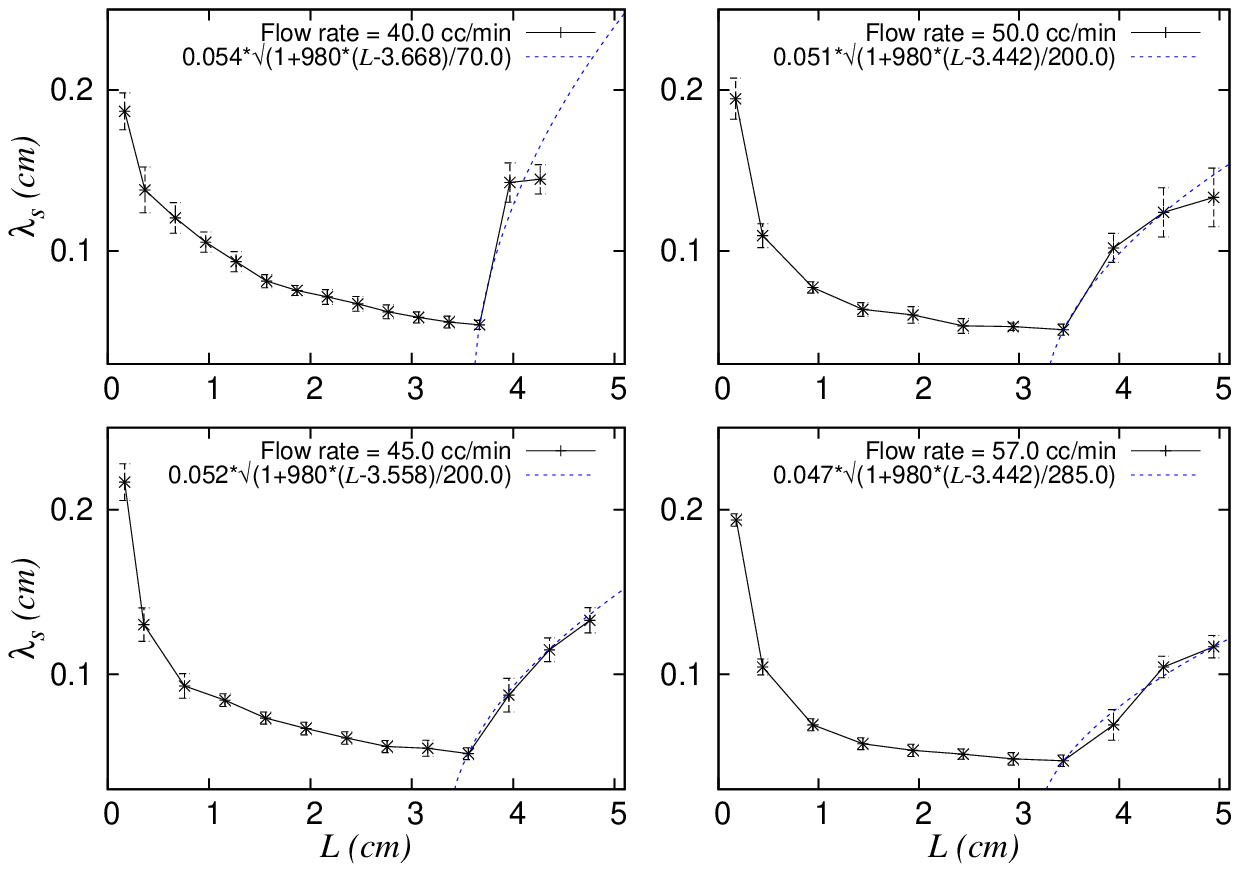}
\caption{Average wavelength ($\lambda_s$) on the water surface of
deep water gravity waves as a function of $L$ for d= 0.95 mm.}
\end{figure}

\begin{figure}
\centering
\includegraphics[width=12cm,height=12cm]{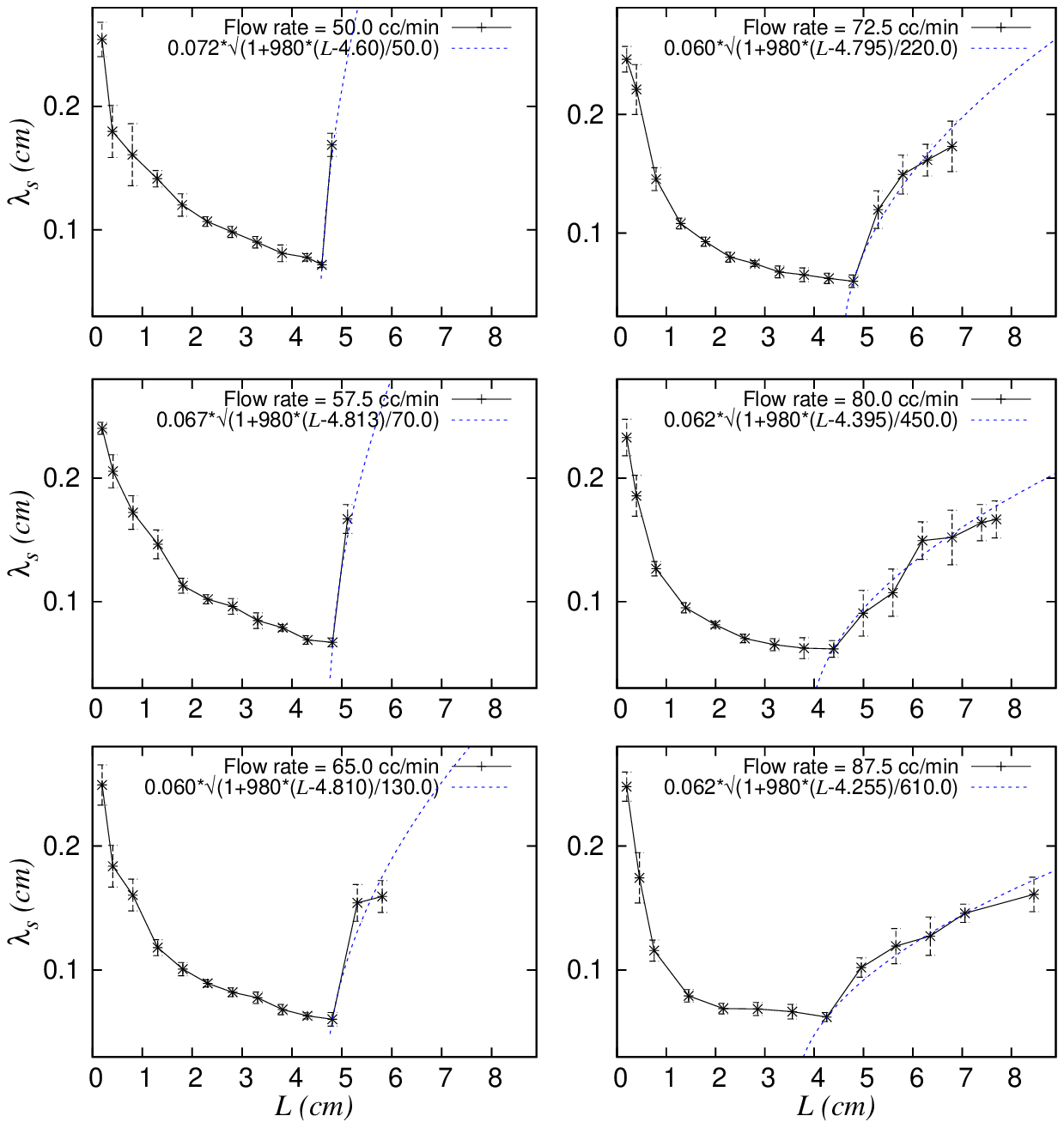}
\caption{Average wavelength ($\lambda_s$) on the water surface of
deep water gravity waves as a function of $L$ for d= 1.26 mm.}
\end{figure}

\begin{figure}
\centering
\includegraphics[width=12cm,height=12cm]{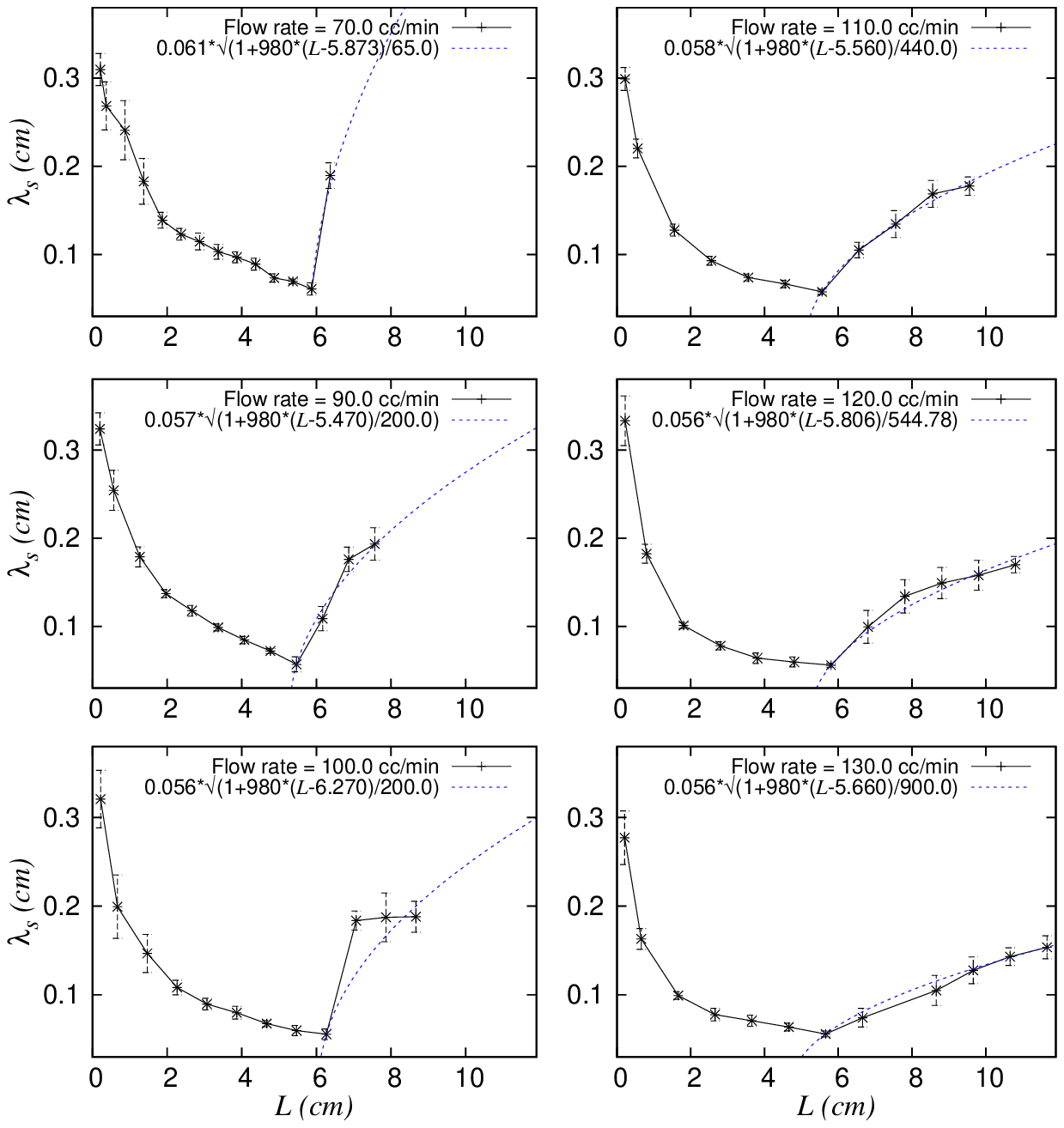}
\caption{Average wavelength ($\lambda_s$) on the water surface of
deep water gravity waves as a function of $L$ for d= 1.54 mm.}
\end{figure}

\begin{figure}
\centering
\includegraphics[width=12cm,height=8cm]{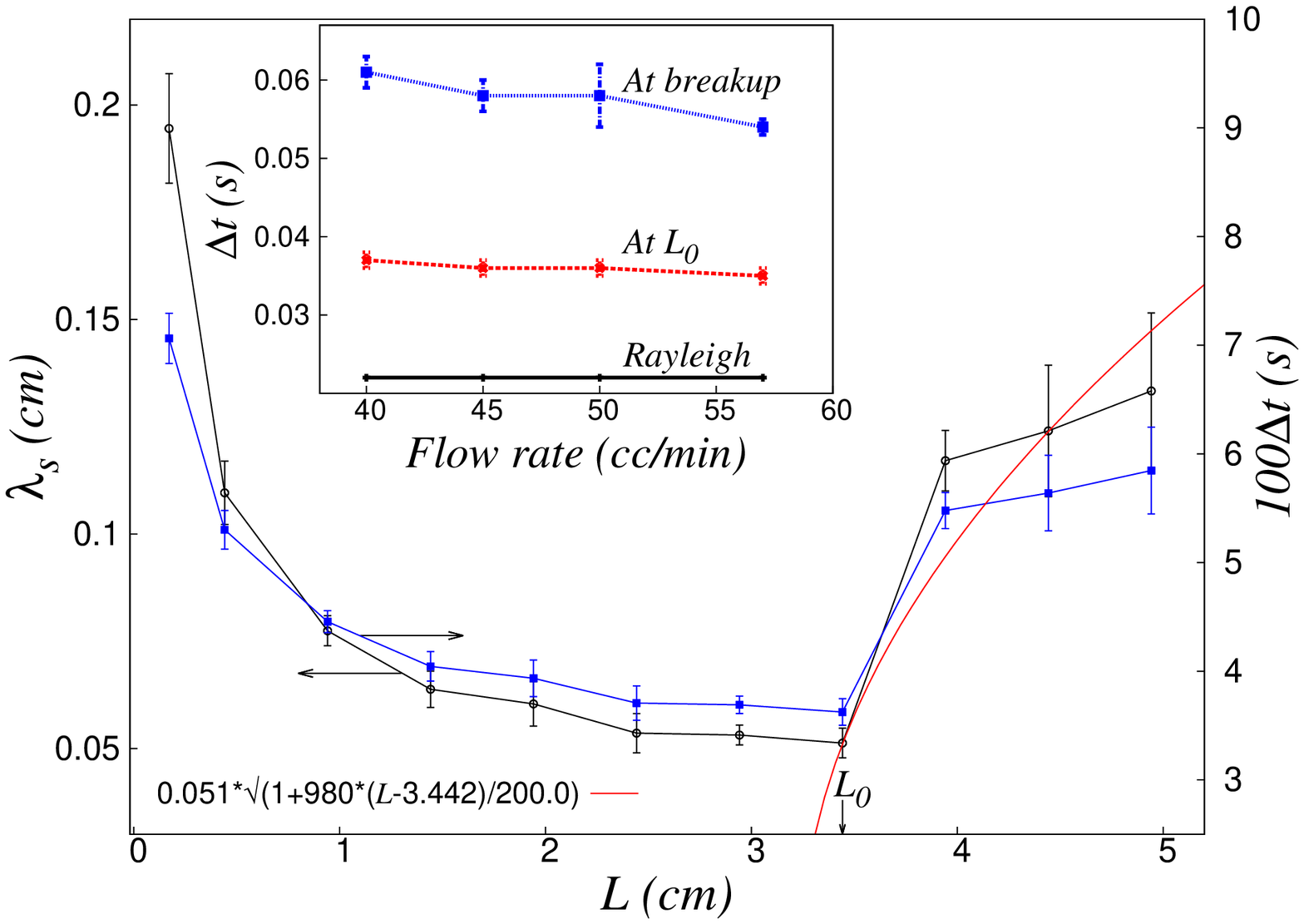}
\caption{Average wavelength ($\lambda_s$) on the water surface and time period
($\Delta$t) of deep water gravity waves as a function of $L$ for d= 0.95 mm and
water flow rate of 50.0 cc/min. A numerical fit to $\lambda_s$ (for $L>L_0$)
is also shown.}
\end{figure}

\begin{figure}
\centering
\includegraphics[width=12cm,height=8cm]{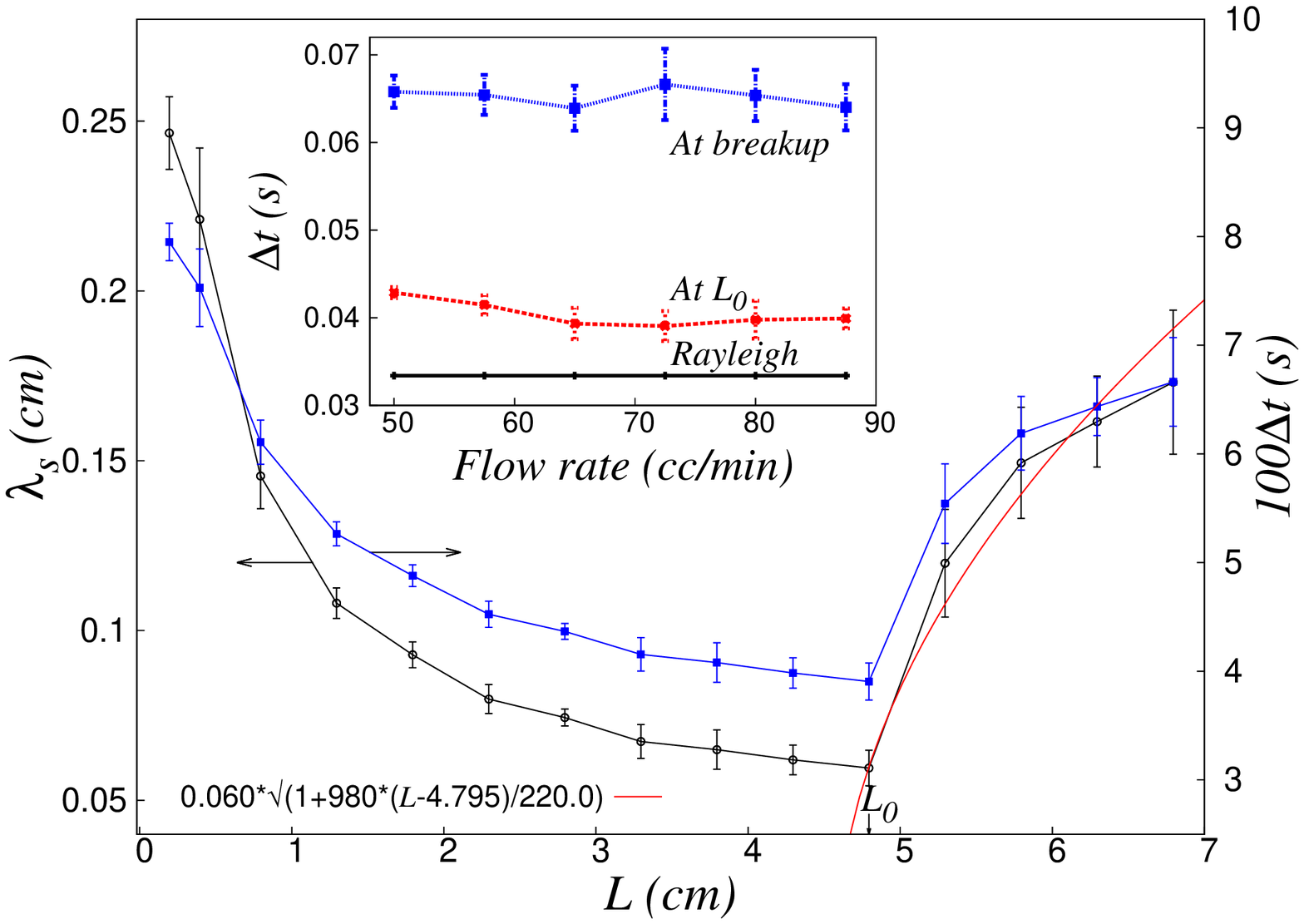}
\caption{Average wavelength ($\lambda_s$) on the water surface and time period
($\Delta$t) of deep water gravity waves as a function of $L$ for d= 1.26 mm and
water flow rate of 72.5 cc/min. A numerical fit to $\lambda_s$ (for $L>L_0$)
is also shown.}
\end{figure}

\begin{figure}
\centering
\includegraphics[width=12cm,height=8cm]{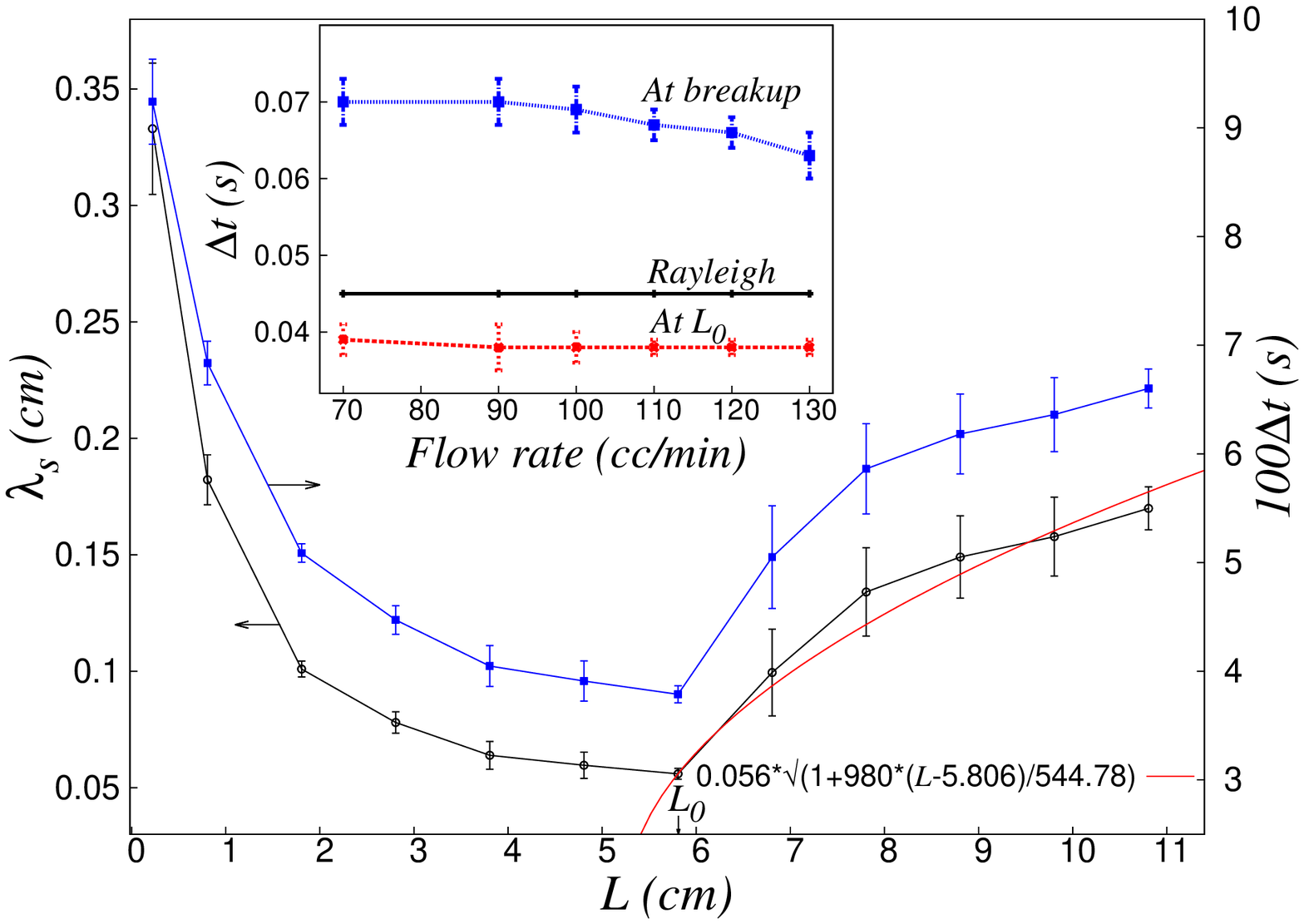}
\caption{Average wavelength ($\lambda_s$) on the water surface and time period
($\Delta$t) of deep water gravity waves as a function of $L$ for d= 1.54 mm and
water flow rate of 120.0 cc/min. A numerical fit to $\lambda_s$ (for $L>L_0$)
is also shown.}
\end{figure}

Figure 2 shows the average number of counts (drops) per second, for a water 
flow rate of 35.0 cc/min and the nozzle inner diameter $d=0.78$ mm, as a 
function of jet-length $l$. The counts range from zero to a saturation value 
and $l_B$ are essentially distributed over a range of values due to the absence
of any fixed external perturbation and presence of unavoidable noise in the 
laboratory as remarked by Donnelly et al \cite{Donnelly}. The mean break-up 
length is thus calculated using the distribution of breakup lengths $l_B$
and it is plotted in Fig. 3.  

The magnified picture of the graph for low values of count rates is shown in 
the inset of Fig. 2. The inset clearly shows that $l_{FB}$ as measured in the 
first set of experiments is larger by about 5 mm compared to $l_{FB}$ measured
in the second set of experiments. Recall that in the first case the recoil 
capillary wave is damped whereas in the latter case it propagates freely up the
jet length. The delay in the process of first breakup of the jet in the first 
set of experiments indicates that the effect of recoil capillary waves do 
exist. Or, equivalently, it shows that the tension of the water surface drags 
the jet down by about 5 mm before it allows the contact between them to 
breakup. Obviously, the effect of recoil capillary wave is small and its mere 
absence cannot delay the breakup indefinitely.

In Fig. 4 the mean $l_B$ and the $l_{FB}$ for the two sets of experiment are 
plotted as a function of water flow rate (or, equivalently, as a function of 
$\sqrt{We}$) for a nozzle of inner diameter 0.78 mm. The difference between 
$l_{FB}$ for the two sets persists for all flow rates. The experiment was 
repeated for various other nozzles with internal diameters, d= 0.69 mm,0.72 mm,
0.82 mm, 0.95 mm, 1.04 mm, 1.14 mm, 1.26 mm, 1.34 mm and 1.54 mm. In all cases 
the results are consistently similar to Fig. 4 and  we arrive at the same 
conclusion about the existence of recoil capillary waves.

As mentioned earlier, Savart's pioneering experiment together with the 
Plateau-Rayleigh theory stimulated further investigations on surface profile 
of the jet and its breakup, for example\cite{Donnelly, Rutland}. We capture 
the waves produced on the surface of the water in the beaker photographically
as the continuous water jet merges into the water body. Figure 5 shows a 
sequence of photographs at different $L=0.168,~0.568,~3.668$, and $4.268$ cms, 
respectively, for nozzle diameter $d=0.95$ mm and a flow rate of 
$FR=$40 cc/min. Similar photpgraphs can be obtained for other nozzle diameters ($d$) 
flow rates $FR$ at different jet length $L$ as well.
Figures 6-8 show representative photographs for $d=0.95$ mm and $FR=50$ cc/min at 
$L=0.942$ cm, $d=1.26$ mm, $FR=72.5$ cc/min at $L=1.295$ cm, and $d=1.54$ mm, 
$FR=120.0$ cc/min at $L=1.806$ cm, respectively. 

 We contend that the waves 
are produced on the water surface due to the time periodic variation of 
cross-section of the jet, that is, due to the periodic crossings of necks and bulges of 
the jet, at the position of the water surface. We vary $L$ from zero till the 
jet-breakup becomes imminent and take phographs of the water surface for 
various $L$. We find that no waves are produced on the water surface when $L$
was zero. On increasing $L$ we could discern the appearance of the circular 
waves for the first time when $L$ was 0.221 cm for $d=1.54$ mm and $FR=120.0$ cc/min. As we gradually lower the 
water surface the circular waves become sharper and then on further increasing 
$L$ the surface waves begin to wane. At a particular $L=L_0$ the waves become 
the least sharp. However, on further lowering the surface, the waves reappear 
with increased sharpness and the waves persist till ultimately the jet breaks 
up before touching the water surface. Figure 5 exhibits
the above mentioned behavior for nozzle diameter $d=0.95$ mm and a flow rate of 
$FR=$40 cc/min. Similar behavior is also observed for other nozzle diameters ($d$) 
flow rates $FR$ at different jet length $L$ as well.

We measure the 
wavelengths $\lambda_s$ of the waves on the water surface using the 
photographs. Figures 9-11 show the measured wavelengths ($\lambda_s$) as a function 
of $L$, respectively, for $d=0.95,~1.26$ and 1.54 mm. The
length scales of the waves are much smaller than the depth ($\approx$ 6 cm) of
the water in the beaker. Thus these waves can be considered as deep water
gravity waves. From the measured $\lambda_s$ we calculate the group velocities
$v_g=\sqrt{\frac{g_0\lambda_s}{8\pi}}$ of these waves, where the acceleration 
due to gravity $g_0$=9.8 m/s$^2$. From these data we calculate the time 
difference ($\Delta t$) between two consecutive crests of the waves. The 
$\Delta t (\propto \sqrt \lambda_s)$, are also plotted in Figs. 12-14 together 
with $\lambda_s$, for the same $d$ values as in Figs. 9-11. This $\Delta t$ 
can also be taken as the time difference between two bulges of the jet 'hitting' 
the water surface.

The measured $\lambda_s$ initially decreases as $L$ increases and reaches a 
minimum value at $L=L_0$ and thereafter, the $\lambda_s$ increases with $L$. 
$L=L_0$ thus marks a sharp boundary in the nature of the water jets. We conjecture 
that the minimum $\lambda_s$ corresponds to the fastest growing Rayleigh 
perturbation in the jet and not the one at the breakup point $L=l_B$. This is 
partially supported by roughly similar values of $\Delta t$. For instance, in the
inset of Fig. 14, $\Delta t\approx 0.038$ s at 
the minimum of $\lambda_s$ (at $L=L_0$) is roughly equal to the time scale 
$\tau=\frac{2\pi}{\omega}\approx 0.045$ s obtained from the estimate of the 
frequency $\omega$ at the maximum of the dispersion curve of the 
Plateau-Rayleigh theory and measured by Goedde and Yuen, (Fig. 7 of 
Ref.\cite{Goedde}). Moreover, the $\Delta t$ at the breakup point ($L=l_B$) of
the jet is comparatively far different from $\tau=0.045$.

The competition between various modes of Rayleigh perturbations makes the 
jet-surface profile variation dynamic before $\lambda_s$ becomes a minimum. In 
other words, the jet-surface profile changes along the length ($L<L_0$) of the 
jet satisfying the stability conditions of Rayleigh theory and affected very 
little by gravity. This is the only way we can explain the variation of 
$\Delta t$, for example in Fig. 14, from about 0.09s to 0.04s keeping the constancy of mean 
mass flow rate of water at any section of the jet.

We contend that for $L<L_0$, the effect of surface tension dominates over the
gravitational effect on the jet whereas at larger $L>L_0$ the gravitational 
effect plays a dominant role making the surface profile of the jet 
nonsinusoidal. The nonsinusoidal surface profile can also be seen from the high 
speed photographs of Ref.\cite{Rutland}. 

In the spirit of Ref.\cite{Scriven}, and considering $\lambda_s$ to be 
proportional to the difference in jet length between two bulgings, we 
numerically fit $\lambda_s(L)$ of the waves on the water surface of the beaker 
(Figs. 9-14) as $\lambda_s(L)=\lambda_s(L_0)\sqrt{1+\frac{2g_0(L-L_0)}{u_0^2}}$ 
taking $u_0$ as a fitting parameter. As can be seen the fit is reasonable. We 
could similarly fit the data for all other nozzles in a range of flow rates. 
Our contention of dominance of gravitational effect for $L>L_0$ thus has good 
experimental support.

\section{Conclusion}
In conclusion, our experiment verifies the existence of recoil capillary waves 
and its effect on jet breakup and points out the relative importance of 
Rayleigh perturbations and gravitational effects on the jet surface profile.
The photographs of surface waves created by the jets on the water surface helps
us measure the wavelegths of the surface waves. The behavior of the surface
wave lengths show a sharp transition as a function of the jet length. The
jet length $L_0$ at the transition point is different for different nozzle
diameters and flow rates. For $L<L_0$ Rayleigh perturbations dominate whereas
for $L>L_0$ gravitaional effect is more important.

\end{document}